\documentclass[fleqn,usenatbib]{mnras}

\usepackage{newtxtext,newtxmath}

\usepackage[T1]{fontenc}

\DeclareRobustCommand{\VAN}[3]{#2}
\let\VANthebibliography\thebibliography
\def\thebibliography{\DeclareRobustCommand{\VAN}[3]{##3}\VANthebibliography}

\usepackage{graphicx}	
\usepackage{amsmath}	
\usepackage{epstopdf}
\usepackage{tabularx}
\usepackage{lineno}
\usepackage{subfigure}

\title[Color evolution of magnetar-powered kilonova]{The color evolution of magnetar-powered kilonova emission in merging neutron star-neutron star systems}

\author[Wang et al.]{
Suo-Ning Wang,$^{1,2}$
Hou-Jun L\"{u},$^{1}$\thanks{E-mail: lhj@gxu.edu.cn (LHJ)}
Xiao-Xuan Liu,$^{1}$
Jared Rice,$^{3}$
Jia Ren,$^{4}$
En-Wei Liang$^{1}$
\\
$^{1}$Guangxi Key Laboratory for Relativistic Astrophysics, School of Physical Science and Technology, Guangxi University, Nanning 530004, China\\
$^{2}$School of Astronomy and Space Science, Nanjing University, Nanjing 210023, China\\
$^{3}$Department of Mathematics and Physical Science, Southwestern Adventist University, Keene, Texas 76059, USA\\
$^{4}$Key Laboratory of Dark Matter and Space Astronomy, Purple Mountain Observatory, Chinese Academy of Sciences, Nanjing 210023, China
}

\date{Accepted XXX. Received YYY; in original form ZZZ}

\pubyear{\the\year{}}

\begin{document}
\label{firstpage}
\pagerange{\pageref{firstpage}--\pageref{lastpage}}
\maketitle

\begin{abstract}
The first direct detection of the gravitational wave (GW) event GW170817 and its electromagnetic (EM) counterpart open a new window for studying of multi-messenger astronomy. However, how to identify the remnant of binary neutron star (NS) merger via EM radiation remain an open question. In this paper, we propose a method of color evolution of kilonova emission to identify its progenitors. We assume that the energy of the kilonova is contributed from radioactive decay, magnetar spin-down, and pulsar wind nebula (PWN). The color evolution of kilonova emission associated with short GRB is significant when the spectrum is thermal emission, while it tends towards a constant when the spectrum is non-thermal radiation. On the other hand, if the central engine is a black hole (BH) which is promptly generated by the NS-NS merger or NS-BH merger, then the kilonova is powered only by the radioactive decay. There is no color evolution at the beginning before the peak of kilonova emission, but is significantly and rapidly increasing after the peak. On the contrary, if the central engine is a magnetar or stable NS, the kilonova emission is contributed from radioactive decay, magnetar, and PWN. The color evolution after the peak of kilonova emission is complex behavior which depends on the rotational energy and spin-down time-scale of magnetar, and finally tend to a constant in the late state.
\end{abstract}

\begin{keywords}
(transients:)  neutron star mergers < Transients
\end{keywords}



\section{Introduction}
\label{section 1}
Short-duration gamma-ray bursts (GRBs) are believed to originate in compact binary mergers, including neutron star-neutron star mergers (NS-NS; \citealt{1986ApJ...308L..43P,1989Natur.340..126E}) and black hole-neutron star mergers (NS-BH; \citealt{1991AcA....41..257P}). GRB 170817A, which is associated with GW 170817 \citep{2017PhRvL.119p1101A}, is the first “smoking gun” evidence in support of the NS-NS merger origin of short-duration GRBs \citep{2017NatAs...1..791C,2017ApJ...848L..14G,2017ApJ...848L..15S,2018NatCo...9..447Z}. On the other hand, several gravitational-wave (GW) events have been reported as NS-BH merger systems, such as GW200115 \citep{2021ApJ...915L...5A}, GW230529 \citep{2023PhRvX..13d1039A}, and a marginal event GW200105 \citep{2021ApJ...915L...5A,2023PhRvX..13d1039A}. However, no electromagnetic (EM) counterpart was found to be accompanied with those events, so NS-BH mergers are still not confirmed as short-duration GRB progenitors \citep{2021ApJ...915L...5A,2023PhRvX..13d1039A}.

Besides short GRBs, kilonova emissions with optical/infrared transients are another electromagnetic (EM) counterpart originating in compact star mergers (\citealt{2019LRR....23....1M} for a review). The first confirmed kilonova emission from a NS-NS merger event is AT2017gfo which was associated with GW 170817 \citep{2017PhRvL.119p1101A,2017Natur.551...80K,2017Sci...358.1565E}. The kilonova emission is generated in the ejected material and is powered by the radioactive decay of r-process nuclei \citep{1998ApJ...507L..59L,2010MNRAS.406.2650M,2011NewAR..55....1B,2011ApJ...732L...6R,2013PhRvD..87b4001H,2013MNRAS.430.2585R,2015NatCo...6.7323Y,2015ApJ...811L..22J,2018ApJ...852L...5M,2025ApJ...988L..46L}. For the convenience of calculation, the ejecta is generally assumed to be nearly-isotropy, but some numerical simulation works of double NS merger suggest that the ejecta or radioactive decay of r-process are more or less an-isotropy \citep{2017PhRvD..95f3016C,2018ApJ...858...52S,2023MNRAS.521.1858C}. Moreover, if the remnant of the NS-NS merger is a newborn magnetar acting as the central engine \citep{2014PhRvD..89d7302L,2015ApJ...805...89L}, the spin energy of the magnetar can power a brighter kilonova called a “merger-nova” \citep{2013ApJ...776L..40Y,2013ApJ...763L..22Z,2014MNRAS.439.3916M,2017ApJ...837...50G,2021ApJ...912...14Y}. On the other hand, the multiband and long-term observations of AT 2017gfo suggest that the appearance of a Pulsar Wind Nebula (PWN) in AT 2017gfo is one possible explanation \citep{2019ApJ...885...60R}. 

How do we identify the NS-NS and NS-BH systems? What is the remnant of binary NS mergers? These remain open questions \citep{2011CRPhy..12..206Z,2014ARA&A..52...43B}. In fact, the most directly method is to adopt GW radiation NS-NS or NS-BH binary system, and constrain the mass range of each object in such binary system \citep{2021ApJ...915L...5A,2023PhRvX..13d1039A}. Another method is to attempt the EM radiation to identify a NS-NS or NS-BH binary system  \citep{2018ApJ...860...62G,2019LRR....23....1M,2020ApJ...889..171K,2020PhR...886....1N,2023Univ....9..245T,2025ApJ...984...77G,2025ApJ...985...42D}. From a theoretical point of view, there are four different types of remnants from binary NS mergers: a black hole, a hypermassive NS supported by differential rotation with a lifetime of about 300 ms, a supramassive NS supported by rigid rotation with a lifetime of hundreds of seconds before collapsing into a BH, and a stable NS \citep{2000A&A...360..171R,2006Sci...311.1127D,2006MNRAS.372L..19F,2010CQGra..27k4105R,2013ApJ...771L..26G,2013ApJ...763L..22Z,2014PhRvD..89d7302L,2014MNRAS.439..744R,2015ApJ...805...89L,2017ApJ...837...50G}. However, identifying the central engine in NS-NS merger remains an open question \citep{2011CRPhy..12..206Z}. There are two ways to answer this question. One is a follow-up to detect post-merger GW emission from the compact star system. If the central engine is a black hole, a weak and high-frequency GW emission can be produced due to the Quasi-normal modes \citep{1999CQGra..16R.159N}. On the other hand, if the central engine is a magnetar it will produce GW radiation, but such a GW signal would be too weak to be detected by the current Advanced LIGO and Advanced Virgo \citep{2017ApJ...835..181L}. Another possibility is the discovery of X-ray internal plateau emission after the short-duration GRBs. The internal plateau emission is believed to be a signature of the supramassive magnetar central engine after the NS-NS merger, but it is not ``smoking gun'' evidence \citep{2006Sci...311.1127D,2008MNRAS.385.1455M,2013MNRAS.430.1061R,2013PhRvD..88f7304F,2013ApJ...763L..22Z,2014MNRAS.441.2433R,2016PhRvD..93d4065G,2017ApJ...849..119C,2020ApJ...898L...6L}.
  
In this work, we propose the method of color index (CI) evolution of the kilonova emission to identify the NS-NS and NS-BH systems and distinguish the remnants in NS-NS mergers. We consider different energy sources, such as radioactive decay of r-process nuclei, named as ``radioactive decay'', rotational energy of magnetar, and the contribution of PWN, and present the color evolution through numerical calculations. Since the temporal evolution of color indices is studied in GRB afterglows \citep{2001A&A...377..450S} and supernovae (SNe) explosions \citep{2004A&A...427..901S}, and color evolution is used to discriminate between various radiation mechanisms and progenitors in GRB and SNe.

This paper is organized as follows. The basic model of kilonova energy sources is shown in \S 2, and \S 3 shows the results of the color evolution of the kilonova emission. Conclusions are drawn in \S 4 with some additional discussion. Throughout the work, we use the notation $Q=10^{n}Q_{n}$ in CGS units (such as $E_{52}=E/(10^{52}~\rm erg)$).

\section{Basic model of kilonova energy sources}
\label{section 2}
\subsection{radioactive decay and magnetar spin-down}
The simplified radiation transfer model, given by \citet{2010ApJ...717..245K} and \citet{2017LRR....20....3M}, is used to describe the emission of the ejecta. Following the method of \citet{2019ApJ...885...60R}, we assume that the ejecta from the NS-NS merger are composed of a series of mass layers with a number $N$, and the velocity of the layers increases from $v_{\rm min}$ (first layer) to $v_{\rm max}$ (last layer) with $N$. Here, we adopt the shell-like profile structure of the ejecta, namely, the entire structure is as a shell. In order to calculate the energy transfer of ejecta, the shell is divided into $N=100$ separated shells. The mass of each layer can be written as $m_{\rm i}=\int_{R_{\rm i}}^{R_{\rm i+1}}4\pi r^{2}\rho_{\mathrm{eje}}(r,t)dr$, and $\rho_{\mathrm{eje}}$ is mass density of ejecta \citep{2014ApJ...784L..28N}
\begin{equation}
    \rho_{\mathrm{eje}}(r,t)=\frac{(\delta-3)M_{\mathrm{eje}}}{4\pi (v_{\mathrm{max}}t)^{3}}\Bigg[\Bigg(\frac{v_{\mathrm{min}}}{v_{\mathrm{max}}}\Bigg)^{3-\delta}-1\Bigg]^{-1}\Bigg(\frac{r}{v_{\mathrm{max}}t}\Bigg)^{-\delta}
\end{equation}
where $M_{\mathrm{eje}}$ is the total mass of the ejecta. The evolution of the thermal energy of the $i$th layer ($E_i$) of the ejecta can be described as
\begin{align}
\frac{dE_{i}}{dt} =\Lambda_{\rm i}\xi L_{\mathrm{sd}}+m_{i}\dot{q}_{\mathrm{r}}\eta_{\mathrm{th}}-\frac{E_{i}}{R_{i}}\frac{dR_{i}}{dt}-L_{i},i=1,...,N.
\end{align}

On the right hand side of Eq.(2), the first term represents the absorbed energy of the $i$th layer from the central engine. In the observer frame, $\Lambda_{\rm i}=(1-e^{-\Delta\tau_{i}})e^{[-(\tau_{\mathrm{tot}}-\tau_{i})]}$. The optical depth of the $i$th layer $\tau_{i}$ can be written as $\tau_{i}=\sum_{i}^{N-1}\Delta\tau_{i},~\mathrm{and}~\Delta\tau_{i}=\int_{R_{i}}^{R_{i+1}}\kappa\rho_{\mathrm{eje}}(r)dr$, where $\kappa$ is the opacity of the ejecta. $\tau_{\mathrm{tot}}=\int_{v_{\mathrm{min}}t}^{v_{\mathrm{max}}t}\kappa\rho_{\mathrm{eje}}(r)dr$ is the total optical depth. $\xi$ is the absorption fraction of $L_{\mathrm{sd}}$, namely, is the fraction of magnetic dipole radiation of the magnetar into non-thermal radiation of PWN which is absorbed by the layers of shell to be a thermal emission, and $L_{\mathrm{sd}}$ can be expressed as \citep{2001ApJ...552L..35Z,2018MNRAS.480.4402L},
\begin{align}
    L_{\mathrm{sd}}(t)=L_{\rm sd, 0}\Bigg(1+\frac{t}{t_{\mathrm{sd}}}\Bigg)^{-2}
\end{align}
\begin{equation}
L_{\rm sd, 0}=\frac{B_{\mathrm{p}}^{2}R^{6}\Omega_{0}^{4}}{6c^{3}}=9.6\times10^{42}R_{6}^{6}B_{\mathrm{p},12}^
{2}P_{0,-3}^{-4}\mathrm{erg~s}^{-1},
\end{equation}
\begin{equation}
    t_{\mathrm{sd}}=\frac{3c^{3}I}{B_{\mathrm{p}}^{2}R^{6}\Omega_{0}^{2}}=2.05\times10^{9}I_{45}B_{\mathrm{p,12}}^{-2}P_{0,-3}^{2}R_{6}^{-6}\text{s},\quad
\end{equation}
where $I$, $R$, $B_{\rm p}$, $P_{\rm 0}$, and $\Omega_{0}$ are the rotational inertia, radius, surface polar magnetic field, initial spin period, and initial angular frequency of the NS, respectively. \textbf{$L_{\rm sd, 0}$} is the initial luminosity, and $t_{\rm sd}$ is the characteristic spin-down timescale. 

The second item on the right side of Eq.(2) is the energy from radioactive decay of heavier elements. The radioactive power per unit mass $\dot{q}_{\mathrm{r}}$ can be written as \citep{2012MNRAS.426.1940K}
\begin{equation}
    \dot{q}_\mathrm{r}=4\times10^{18}\biggl[\frac{1}{2}-\frac{1}{\pi}\arctan\biggl(\frac{t-t_0}{\sigma}\biggr)\biggr]^{1.3}\text{erg s}^{-1}\text{g}^{-1}.
\end{equation}
Here, we adopt $t_0=1.3~{\rm s}$ and $\sigma=0.11~{\rm s}$ \citep{2012MNRAS.426.1940K}. The thermalization efficiency of the radioactive power reads as \citep{2016ApJ...829..110B,2017LRR....20....3M,2019ApJ...885...60R}
\begin{equation}
\eta_{\mathrm{th}}=0.36\biggl[\exp(-0.56t_{\mathrm{day}})+
    \frac{\ln(1+0.34t_{\mathrm{day}}^{0.74})}{0.34t_{\mathrm{day}}^{0.74}}\biggr],
\end{equation}
where $t_{\mathrm{day}}$ is the time in units of days. The third and fourth terms of the right side in Eq.(2) are cooling with adiabatic expansion and radiation, respectively. The observed luminosity from the $i$th layer $L_i$ can be estimated as
\begin{equation}
    L_{i}=\frac{E_{i}}{\max\left\{t_{\mathrm{d}}^{i},t_{\mathrm{lc}}^{i}\right\}},
\end{equation}
where $t_{\mathrm{lc}}^{i}=R_{i}/c$ is the light-crossing time scale, and $t_\mathrm{d}^i$ is the photon diffusion timescale which can be read as
\begin{equation}
    t_\mathrm{d}^i\simeq\frac{\kappa}{4\pi R_ic}\sum_{j=i}^{N-1}m_j,
\end{equation}

Finally, by summing the radiation luminosity from all of the layers, one can get the total bolometric luminosity, i.e.,
\begin{equation}
    L_{\mathrm{e}}=\sum_{i=1}^{N-1}L_i.
\end{equation}
The radiation of the ejecta emanates from the photosphere $R_{\rm ph}$ with a blackbody radiation spectrum. By setting $\tau_{\mathrm{ph}}=\int_{R_{\mathrm{ph}}}^{R_{\mathrm{max}}}\rho_{\mathrm{eje}}(r)dr=1$ for $\tau_{\mathrm{tot}}>1$, and $R_{\mathrm{ph}}=R_{\mathrm{min}}$ for $\tau_{\mathrm{tot}}<1$, the effective temperature can be defined as \citep{2018ApJ...861..114Y}
\begin{equation}
    T_{\mathrm{e}}=\left(\frac{L_{\mathrm{e}}}{4\pi\sigma_{\mathrm{SB}}R_{\mathrm{ph}}^{2}}\right)^{1/4},
\end{equation}
where $\sigma_{\mathrm{SB}}$ is the Stephan-Boltzman constant. For a given frequency $\nu$, the observed flux density can be calculated as
\begin{equation}
    F_{\nu}(t)=\frac{2\pi h\nu^{3}}{c^{2}}\frac{1}{\exp(h\nu/kT_{\mathrm{e}})-1}\frac{R_{\mathrm{ph}}^{2}}{D_{L}^{2}},
\end{equation}
where $h$ and $k$ are the Planck constant and Boltzmann constant, respectively. $D_{\rm L}$ is the luminosity distance of the source.


\subsection{Pulsar wind nebula}
A magnetar may be formed as the central engine after a NS-NS merger, and the pulsar wind can interact with the ejecta to produce both forward shock (between the shocked and unshocked ejecta) and reverse shock (between the shocked and unshocked pulsar wind). The pulsar wind nebula (PWN) is composed of the shocked material between the forward and reverse shocks \citep{1998PhRvL..81.4301D,2006ARA&A..44...17G,2019ApJ...885...60R,2024ApJ...963..156W}. If the pulsar wind is constrained by the surrounding medium, it will create a “termination shock”. If this is the case, electrons and positrons can be accelerated to ultra-relativistic energies, and the magnetic field can be amplified. The magnetic energy density $U_{\rm PWN}$ can be described as \citep{2010ApJ...715.1248T,2013MNRAS.429.2945T,2016MNRAS.461.1498M}
\begin{equation}
    U_{\mathrm{PWN}}=\frac{B_{\mathrm{PWN}}^{2}}{8\pi}=\frac{3}{4\pi}\epsilon_{B}R_{\mathrm{PWN}}^{-3}(t)\int_{0}^{t}L_{\mathrm{sd}}(t)dt,
\end{equation}
where $R_{\mathrm{PWN}}=[3M_{\mathrm{eje}}/(4\pi nm_{p})]^{1/3}$ is the radius that the accelerated leptons and the amplified magnetic fields fill the PWN, and $n$ and $m_{\rm p}$ are the number medium density and the proton mass, respectively. $\epsilon_{B}$ is the fraction of the magnetic energy density to the total energy density. An electron spectrum with a broken power-law is adopted to do the calculations \citep{2015ApJ...805...82M},
\begin{equation}
    \frac{d\dot{n}_e}{d\gamma_e}\propto\begin{cases}\gamma_e^{-q_1},&\gamma_m\leqslant\gamma_e<\gamma_b,\\\gamma_e^{-q_2},&\gamma_b\leqslant\gamma_e\leqslant\gamma_M,\end{cases}
\end{equation}
where $q_{1}\sim1-2$ and $q_{2}\sim2-3$ are the low and high energy spectral index, respectively. $\gamma_{b}\sim10^{4}-10^{6}$ is the break Lorentz factor, and $\gamma_m$ and $\gamma_M$ are the minimum and maximum Lorentz factor of leptons, respectively. 

For the synchrotron radiation of the PWN, the characteristic synchrotron frequency $\nu_{b}$ and the cooling frequency $\nu_{c}$ can be expressed as
\begin{equation}
    \nu_{b}\approx\frac{3}{4\pi}\gamma_{b}^{2}\frac{q_{e}B_{\mathrm{PWN}}}{m_{e}c},
\end{equation}
\begin{equation}
    \nu_{c}\approx\frac{3}{4\pi}\gamma_{c}^{2}\frac{q_{e}B_{\mathrm{PWN}}}{m_{e}c},
\end{equation}
where $q_{e}$ is the charge of the electron, $\gamma_{c}=6\pi m_{e}c/(\sigma_{\mathrm{T}}B_{\mathrm{PWN}}^{2}t)$ is the cooling Lorentz factor, and $\sigma_{\mathrm{T}}$ is the Thomson cross-section \citep{1998ApJ...497L..17S}. By considering a relativistic shock propagating through a uniform cold medium, the shock undergoes adiabatic and radiative hydrodynamic evolution. Within this scenario, the radiated spectrum can be divided into two different regions, e.g., fast-cooling ($\nu_{c}<\nu_{b}$) and slow-cooling ($\nu_{c}>\nu_{b}$) cases \citep{2016MNRAS.461.1498M,2019ApJ...885...60R}.

(1) Fast-cooling regime:
\begin{equation}
\begin{aligned}
&\nu L_{\nu}^{\mathrm{PWN}}= \begin{cases}A\xi L_{\mathrm{sd}}(\nu_c/\nu_b)^{(2-q_1)/2}(\nu/\nu_c)^{(3-q_1)/2},&\nu\leqslant\nu_c,\\A\xi L_{\mathrm{sd}}(\nu/\nu_b)^{(2q_1)/2},&\nu_c\leqslant\nu\leqslant\nu_b,\\A\xi L_{\mathrm{sd}}(\nu/\nu_b)^{(2-q_2)/2},&\nu_b\leqslant\nu\leqslant\nu_M;\end{cases}
\end{aligned}
\end{equation}

(2) Slow-cooling regime:
\begin{equation}
\begin{aligned}
&\nu L_{\nu}^{\mathrm{PWN}}= 
\begin{cases}A \xi L_{\mathrm{sd}}(\nu_b/\nu_c)^{(3-q_2)/2}(\nu/\nu_b)^{(3-q_1)/2},&\nu\leqslant\nu_b,\\A \xi L_{\mathrm{sd}}(\nu/\nu_c)^{(3-q_2)/2},&\nu_b\leqslant\nu\leqslant\nu_c,\\A \xi L_{\mathrm{sd}}(\nu/\nu_c)^{(2-q_2)/2},&\nu_c\leqslant\nu\leqslant\nu_M,\end{cases}
\end{aligned}
\end{equation}
Here, $A=\frac{(2-q_1)(q_2-2)}{2(q_2-q_1)}$ is a constant dependent on $q_1$ and $q_2$, and the absorption fraction $\xi=\eta\epsilon_{e}=\eta(1-\epsilon_{B})$ with $\eta=\min\{1,\:(\nu_b/\nu_c)^{(q_2-2)/2}\}$ is adopted in our calculations \citep{2006MNRAS.369..197F}.

Based on the above equations, one has $\int_0^{\infty}L_{\nu}d\nu\approx\xi L_{\mathrm{sd}}$. In our calculations, we ignore the inverse Compton scattering process because it is not the main focus of this work. Moreover, we also assume that the radiation spectrum does not change when the photons pass through the ejecta. If this is the case, the observed flux from PWN can be written as
\begin{equation}
    F_{\nu}^{\mathrm{PWN}}=\frac{L_{\nu}e^{-\tau_{\mathrm{tot}}}}{4\pi D_{L}^{2}}.
\end{equation}
The observed kilonova emission flux is contributed to by the radioactive decay and spin-down energy of magnetar ($F_{\nu}$) and $F_{\nu}^{\mathrm{PWN}}$ from the PWN, i.e.
\begin{equation}
    F_\nu^{\text{tot}}=F_\nu+F_\nu^{\text{PWN}}.
\end{equation}
Thus, we can determine the monochromatic magnitude of the emission of the kilonova emission \citep{2018ApJ...861..114Y}.
\begin{equation}
    M_{\nu}=-2.5\log_{10}\frac{F_{\nu}^{\mathrm{tot}}}{3631}\mathrm{Jy}.
\end{equation}

\begin{figure*}
    \vspace{2cm}
    \centering
    \subfigbottomskip=20pt
    \subfigcapskip=2pt
    \subfigure{\includegraphics[angle=0,scale=1]{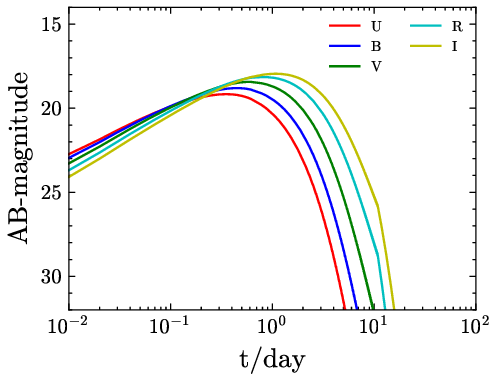}}
    \quad  
    \subfigure{\includegraphics[angle=0,scale=1]{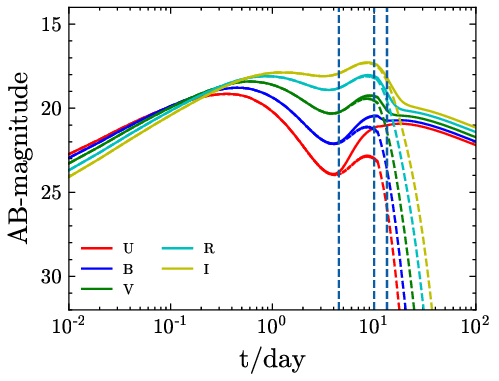}}
\caption{Light curves of kilonova emission in U-, R-, B-, I-, and V-band by only considering radioactive decay (left panel), radioactive decay $+$ magnetar spin-down (dashed lines in right panel), and radioactive decay $+$ magnetar spin-down $+$ PWN (solid lines in right panel). The first blue vertical dashed line indicates that the outermost optical depth is less than one, and we can see the contribution of the inner layer. The second vertical line represents the moment of maximum brightness. The third vertical line represents the beginning of PWN dominant phase.}
\label{fig:1}
\end{figure*}

Based on the above equations, one can calculate the kilonova emission for different energy sources. Figure \ref{fig:1} shows the results of numerical calculations of the kilonova emission in the U-, R-, B-, I-, and V-bands by considering only radioactive decay (left panel), radioactive decay$+$magnetar spin-down (dashed lines in the right panel), and radioactive decay$+$magnetar spin-down$+$PWN (solid lines in the right panel). The parameters are taken as $L_{\mathrm{sd},0}=10^{42}~{\rm erg~s^{-1}}$, $v_{\rm min}=0.1c$, $v_{\rm max}=0.3c$, $\delta=2$, $M_{\mathrm{eje}}=0. 03M_{\odot}$, $\kappa=5\mathrm{~g~cm}^{-2}$, $t_{\mathrm{sd}}=10^{6}\:\mathrm{s}$, $\epsilon_{B}=0. 01$, $\gamma_{b}=10^{4}$, $\gamma_{M}=10^{6}$, $q_1=1.8$, $q_2=2.2$, and luminosity distance $D_{\rm L}=40~{\rm Mpc}$. We find that the radioactive decay dominates the early stages and begins to decay after reaching its peak about one day. Compared to the radioactive decay, the added magnetar spin-down energy dominates the contribution to the kilonova emission after several days. In this case, the optical depth of the outermost layer is less than one, and the contribution from the inner layer can be gradually seen. It means that the rotational energy mainly affects the inner layers of the ejecta, and the duration of the influence mainly depends on the opacity of the ejecta rather than the spin-down timescale. The spin-down timescale and the initial luminosity of the magnetar mainly affect the brightness of the second peak of the kilonova emission in Figure \ref{fig:1}. Furthermore, it is found that a significant contribution to the kilonova emission of the added PWN occurs at a later time by comparing radioactive decay$+$magnetar spin-down.

\section{Color evolution of kilonova emission}
\label{section 3} 
The color index is defined as the difference in magnitude between any two filters of the kilonova emission, and the color evolution is that the color index changes with time. Based on the definition, the color index can be written as
\begin{equation}
\mathrm{CI}=\mathrm{m}(\lambda_2)-\mathrm{m}(\lambda_1),
\end{equation}
where $\lambda_{1}$ and $\lambda_{2}$ are the wavelengths of two bands. If we set $\lambda_{2}<\lambda_{1}$, the smaller (or larger) value of CI, the hotter (or cooler) the kilonova. In this work, we focus on the variation of CI between two adjacent bands, e.g. U-B, B-V, V-R, and R-I.

\begin{figure*}
    \vspace{0cm}
    \centering
    \subfigbottomskip=20pt
    \subfigcapskip=2pt
    \subfigure{\includegraphics[angle=0,scale=1]{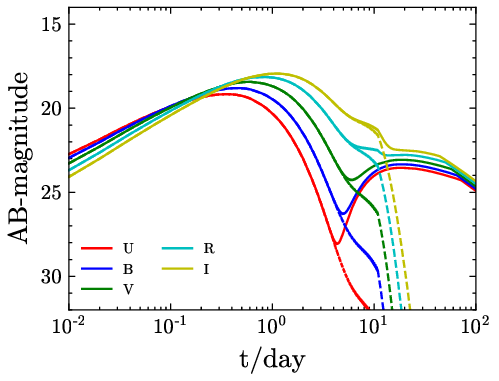}}
    \quad  
    \subfigure{\includegraphics[angle=0,scale=1]{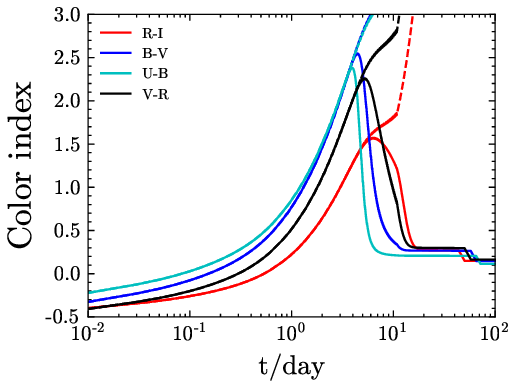}}
    \\
    \subfigure{\includegraphics[angle=0,scale=1]{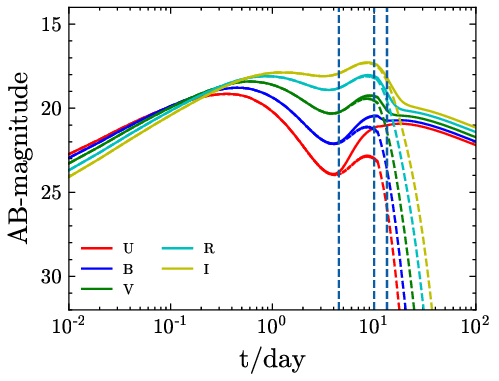}}
    \quad  
    \subfigure{\includegraphics[angle=0,scale=1]{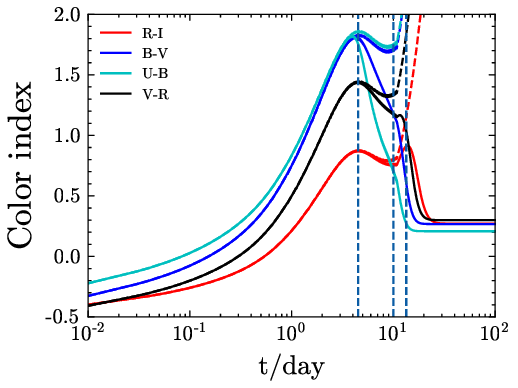}}
     \\
    \subfigure{\includegraphics[angle=0,scale=1]{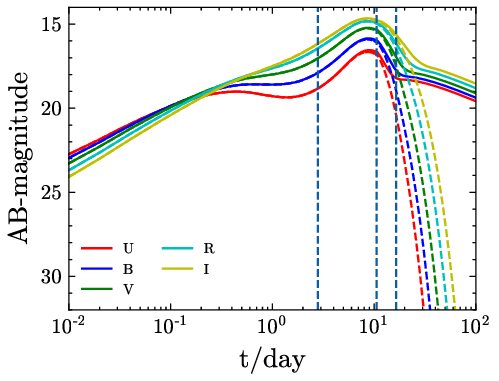}}
    \quad  
    \subfigure{\includegraphics[angle=0,scale=1]{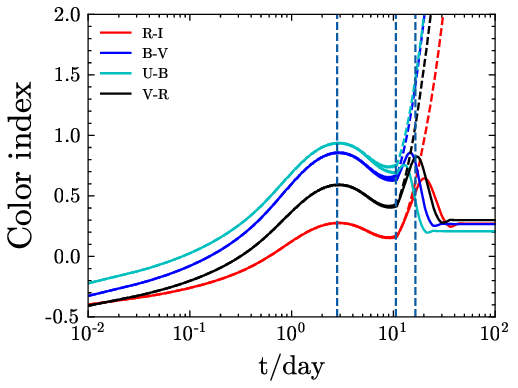}}
\caption{Light curves (left panels) and color evolution (right panels) of kilonova emission is generated by NS-NS merger by considering both radioactive decay $+$ magnetar spin-down (dashed lines) and radioactive decay $+$ magnetar spin-down $+$ PWN (solid lines). For fixed $t_{\mathrm{sd}}=10^{6}~{\rm s}$, the magnetar spin-down energy injection is adopted as $L_{\mathrm{sd},0}=10^{41}~{\rm erg~s^{-1}}$ (top panels), $~10^{42}~{\rm erg~s^{-1}}$ (middle panels), $~10^{43}~{\rm erg~s^{-1}}$ (bottom panels), respectively.}
\label{fig:2}
\end{figure*}

By assuming that the kilonova emission is powered by a NS-NS merger with a magnetar central engine, one can present the color evolution of the kilonova emission by considering both radioactive decay$+$magnetar spin-down and radioactive decay$+$magnetar spin-down$+$PWN for numerical calculation. The left panels in Figure \ref{fig:2} show the calculated kilonova light curves, which are similar to the right panel in Figure \ref{fig:1} with different spin-down luminosities (e.g., $L_{\rm sd,0}=10^{41}$, $10^{42}$, and $10^{43}\rm ~erg~s^{-1}$). It is found that the energy injection from the magnetar has a significant effect on the light curve of the kilonova emission, namely, the larger spin-down luminosity, the brighter kilonova emission. It seems to be two peaks of kilonova emission by considering radioactive decay $+$ magnetar spin-down $+$ PWN, and the first peak of kilonova emission is contributed from the radioactive decay in the early stage. During this process, the optical depth is high enough. So that, the PWN emission is mainly absorbed by the inner region of the ejecta, and the radiation cooling of the ejecta mainly occurs in the outer region. When the total optical depth is decreased, the energy deposited inside is released, and the second peak appears at the total optical depth close to one. The second peak is contributed by the spin-down energy of the magnetar, and the late stage of the kilonova emission is dominated by the contribution of the PWN.

On the other hand, we also calculate the corresponding color evolution of the kilonova emission in the right panels of Figure \ref{fig:2}. (1). By considering only the radioactive decay of kilonova emission: we find that there is no significant color evolution at the beginning before the first peak of kilonova emission due to the high temperature, but is significantly and rapidly increasing when the temperature drops into the optical band after the peak. (2). By considering the energy sources from the radioactive decay$+$ magnetar spin-down, it is found that the color evolution is continuously increased if the contribution of the magnetar spin-down energy of the kilonova is much smaller than that of the radioactive decay (the top right panel of Figure \ref{fig:2}). On the contrary, if the contribution of the magnetar spin-down energy of the kilonova is comparable to or larger than that of the radioactive decay, the color evolution decreases and then increases (the middle and bottom panels on the right of Figure \ref{fig:2}). (3). By considering the energy sources of the kilonova emission for radioactive decay$+$ magnetar spin-down $+$ PWN, the color evolution at the late state is rapidly decreasing and then tend to a constant. It is clear to see that the behavior of the color evolution of the kilonova emission by considering the radioactive decay $+$ magnetar spin-down is quite different from that of considering the radioactive decay $+$ magnetar spin-down $+$ PWN. An interesting possibility is that it can be used to identify the central engine (BH or magnetar) from the remnant of NS-NS mergers.

\begin{figure*}
    \vspace{0cm}
    \centering
    \subfigbottomskip=20pt
    \subfigcapskip=2pt
    \subfigure{\includegraphics[angle=0,scale=1]{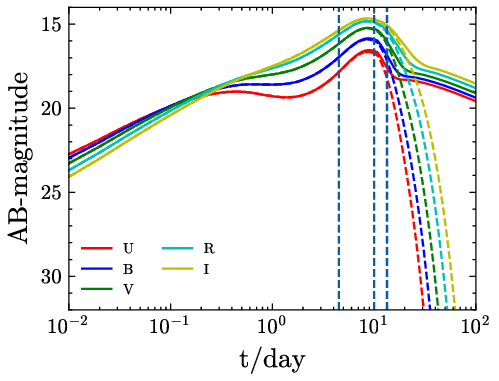}}
    \quad  
    \subfigure{\includegraphics[angle=0,scale=1]{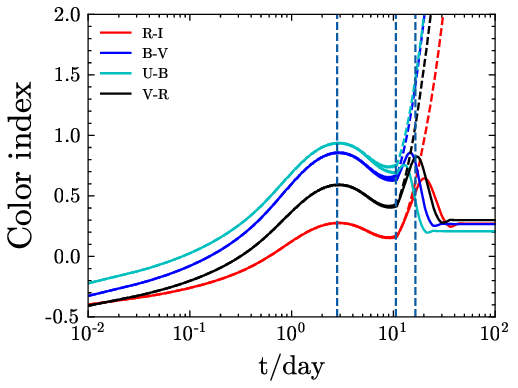}}
    \\
    \subfigure{\includegraphics[angle=0,scale=1]{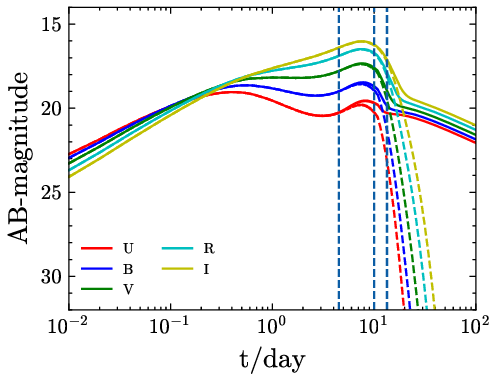}}
    \quad  
    \subfigure{\includegraphics[angle=0,scale=1]{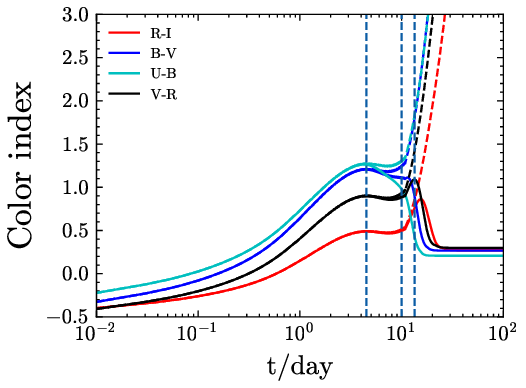}}
     \\
    \subfigure{\includegraphics[angle=0,scale=1]{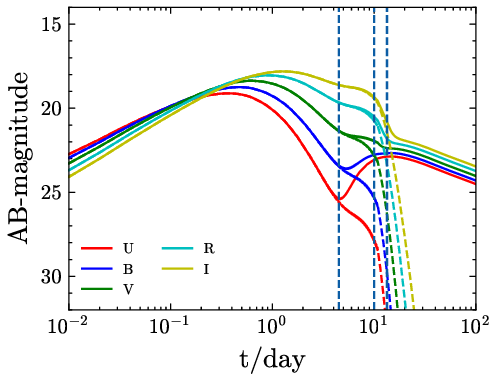}}
    \quad  
    \subfigure{\includegraphics[angle=0,scale=1]{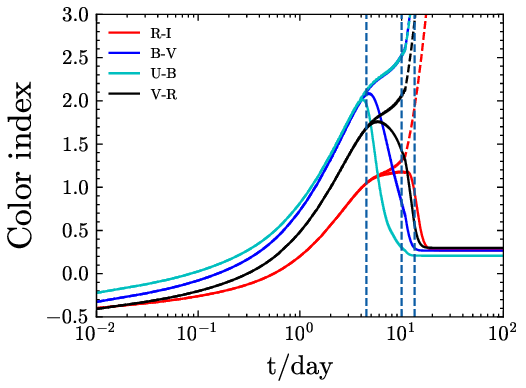}}
\caption{Similar to Figure \ref{fig:2}, but fixed $L_{\mathrm{sd},0}=10^{43}~{\rm erg~s^{-1}}$ with varying magnetar spin-down timescale as $t_{\mathrm{sd}}=10^{6}~{\rm s}$ (top panels), $~10^{5}~{\rm s}$ (middle panels), $~10^{4}~{\rm s}$ (bottom panels), respectively.}
\label{fig:3}
\end{figure*}

Moreover, in order to test how sensitive kilonova emission is dependent on the spin-down timescale by considering both radioactive decay$+$magnetar spin-down and radioactive decay$+$magnetar spin-down$+$PWN, we calculate the kilonova emission and its color evolution by adopting different values of opacity and spin-down timescale. Figure \ref{fig:3} shows the kilonova emission and its color evolution with varying spin-down timescale (e.g., $t_{\rm sd} = 10^6$ s, $10^5$ s, and $10^4$ s) for fixed $L_{\rm sd,0}=10^{43}\rm ~erg~s^{-1}$. It is found that the brightness of kilonova emission is decreasing for decreased $t_{\rm sd}$. The color evolution of the kilonova emission in the early stage is increasing when $t_{\rm sd}$ is decreased, but it seems to be not significantly dependent on the values of $t_{\rm sd}$ in the late stage for PWN.

\begin{figure*}
 \includegraphics [angle=0,scale=1] {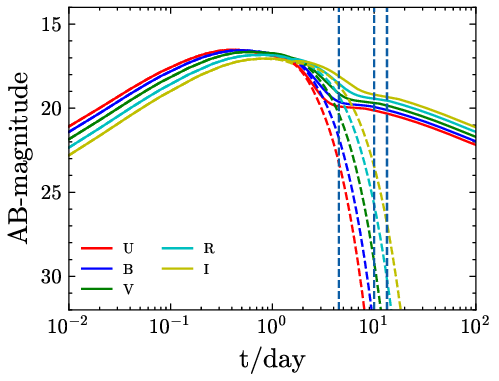}
 \includegraphics [angle=0,scale=1] {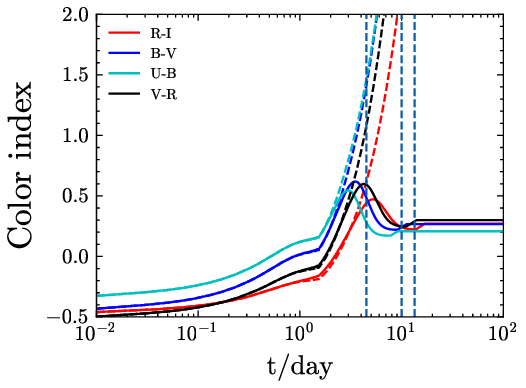}
 \includegraphics [angle=0,scale=1] {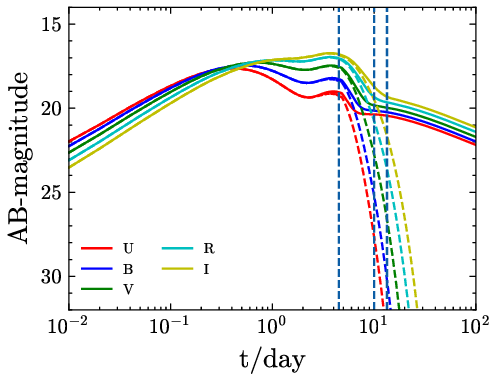}
 \includegraphics [angle=0,scale=1] {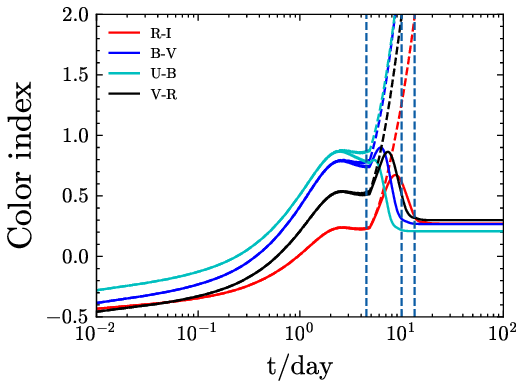}
  \includegraphics [angle=0,scale=1] {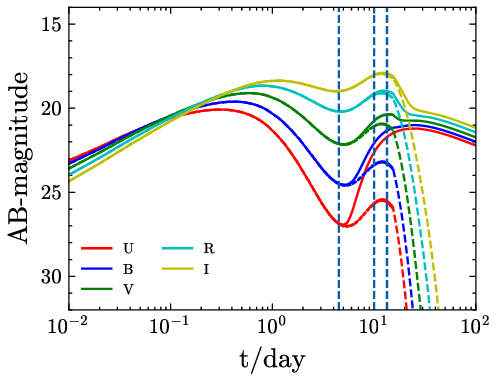}
 \includegraphics [angle=0,scale=1] {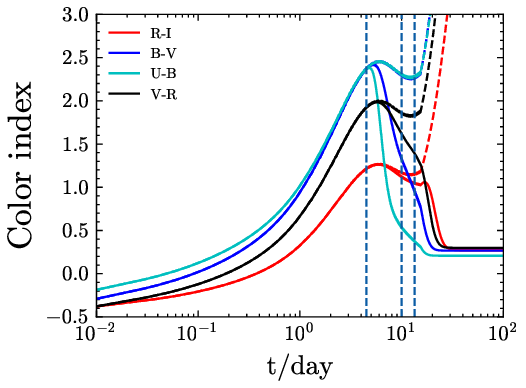}
 \centering
 \caption{Kilonova light curves that are similar to Figure \ref{fig:2}, but with the values of the parameters $L_{\mathrm{sd},0}=10^{42}~{\rm erg~s^{-1}}$ and  $t_{\rm sd} = 10^6$ s. The opacity is adopted as $\kappa=0.1 \mathrm{~cm}^{2}\mathrm{~g}^{-1}$ (top panels), $~1~\mathrm{~cm}^{2}\mathrm{~g}^{-1}$ (middle panels), $~10 \mathrm{~cm}^{2}\mathrm{~g}^{-1}$ (bottom panels), respectively.}
 \label{fig:4}
\end{figure*}

Figure \ref{fig:4} shows the kilonova emission and its color evolution with varying opacity (e.g., $\kappa=0.1\rm~ cm^{2}~g^{-1}$, $1\rm~ cm^{2}~g^{-1}$, and $5\rm~ cm^{2}~g^{-1}$) for fixed $L_{\rm sd,0}=10^{42}\rm ~erg~s^{-1}$ and $t_{\rm sd} = 10^6$ s. It is found that the brightness of kilonova emission (without PWN) is decreasing when opacity $\kappa$ is increasing, but the contribution from PWN of kilonova emission increases. If this is the case, the color evolution of the kilonova emission in the early stage is also increasing when the opacity $\kappa$ is increased, but it also seems to be not significantly dependent on the values of opacity $\kappa$ in the late stage for PWN. Moreover, the behavior of kilonova and its color evolution maybe also depend on ejecta geometry and viewing angle, and they require convincing numerical simulations which are beyond the scope of this paper.

\section{Detection prospects of the color evolution of kilonova emission}
\label{section 4} 
The color evolution of kilonova emission is an important probe for evaluating the post-merger remnants which result from mergers of binary compact objects. One question is whether such features of color evolution can be observed by current or coming optical telescopes. In this section, we will explore the detectability of such kilonova emission and its color evolution by invoking several optical telescopes, such as Zwicky Transient Facility (ZTF), Nancy Grace Roman Space Telescope (Roman), Legacy Survey of Space and Time (LSST), and James Webb Space Telescope (JWST).

By adopting $L_{\mathrm{sd},0}=10^{41}~{\rm erg~s^{-1}}$, $t_{\rm sd} = 10^6$ s, $\kappa=5 \mathrm{~cm}^{2}\mathrm{~g}^{-1}$, and the luminosity distance at $D_{\rm L}=100$ Mpc which corresponds to the distance of typical gravitational wave (GW) detection horizon, we calculate the brightness of multi-bands of possible kilonova emission and its color evolution, and then compare with that of detection thresholds of ZTF, Roman, LSST, and JWST. We find that the calculated lowest magnitudes within 100 days for U-, B-, V-, R-, and I-band are 30, 28, 26, 26, and 26 magnitudes, respectively. For ZTF, the maximum stacking depth is about 21.5-22 magnitude, and it can capture bright optical transients well, but is not suitable for weak kilonovae. For LSST, single exposure limit is about 24.5 magnitude, and the multi band cumulative limits can reach at the depth of 26-28 magnitude. If this is the case, one can be observed for all bands excepting for the U-band. For JWST, it has a near infrared limit of 28 magnitude, and can achieve a deep field up to 30 magnitude in the near-infrared band which is sufficient to observe such kilonova emission. For the upcoming Roman, the single exposure depth in the near-infrared band is around 26.5 magnitude, with a maximum depth of around 28-29 magnitude. Thus, this kinds of faint kilonovae are possible detectable by advanced telescopes like LSST, JWST, and Roman.

\section{Conclusion and discussion}
\label{section 5} 
The merger of binary neutron stars is a potential source of gravitational waves, short GRBs, and kilonova emission (\citealt{2014ARA&A..52...43B} for a review). The kilonova emission is generated from the ejected material and powered by radioactive decay of r-process nuclei \citep{1998ApJ...507L..59L,2010MNRAS.406.2650M,2011NewAR..55....1B}. The possible remnant of such a merger may be a stellar-mass BH, a rapidly spinning supramassive magnetar, or a stable NS \citep{2016PhRvD..93d4065G}. The observed X-ray internal plateau emission in some short GRBs, such as GRB 160821B \citep{2017ApJ...835..181L}, and the X-ray flare emission from event of GW170817/GRB 170817A \citep{2017PhRvL.119p1101A}, suggest that a long-lived magnetized NS may survive after the NS-NS merger \citep{2019MNRAS.483.1912P,2019MNRAS.486.4479L}. If this is the case, then the main power of the kilonova emission from a double NS merger should no longer be limited to the radioactive decay, but include the contribution from the rotational energy of the magnetar and the PWN \citep{2019ApJ...885...60R}.

In this work, we adopt the numerical calculations to simulate the light curves and its color evolution of kilonova emission generated by NS-NS merger by considering both radioactive decay $+$ magnetar spin-down and radioactive decay $+$ magnetar spin-down $+$ PWN. We find that the color evolution of the kilonova emission depends on the spin-down luminosity of the magnetar and spin-down time-scale in the early stage. Some interesting results are summarized as following:
\begin{itemize}
 \item If one can observe the kilonova emission and the color index evolution of the PWN, such as the kilonova emission exhibits a clearly bimodal behavior with an increasing or decreasing in color evolution, and then tends to a constant at the later stage, one possible explanation is the magnetar central engine which is the remnant from the NS-NS merger.
 \item If the central engine is a BH which is the prompt remnant from the NS-NS merger or NS-BH merger, and the kilonova emission is driven only by the radioactive decay. If this is the case, there is no color evolution at the beginning before the peak of kilonova emission, but is significantly and rapidly increasing after the peak. However, the method of color evolution cannot be used to distinguish the formation channel of BH, namely, whether the BH is from mergers of compact star (e.g., NS-NS and NS-BH) or the collapse of supramassive NS.
 \item If the central engine is a magnetar or a stable NS from NS-NS merger, the kilonova emission is contributed from radioactive decay of r-process nuclei, magnetar, and PWN. The color evolution after the peak of kilonova emission is complex behavior which depends on the rotational energy and spin-down time-scale of magnetar, and finally tend to constant in the late state.
\end{itemize}

Based on Section 2, it is clear to see that the kilonova emission from radioactive decay $+$ magnetar spin-down is typical thermal radiation, while the emission from PWN is synchrotron radiation. It means that we can identify the kilonova emission radiation mechanisms only based on the behavior of the color evolution. From the observational point of view, it is difficult to observe multi-band and having enough data from the kilonova emission unless the distance to the kilonova is small enough or the kilonova is bright enough.

However, this method has a significant limitation to compare with observations. A color evolution of kilonova emission can occur at approximately one week, after which more observational data are required to make judgment. Moreover, the color evolution of kilonova emission tends to become constant after 10 days or longer, which is a great challenge for current observations. In general, the kilonova emission is too weak to be detected after ten days unless the distance is nearby enough or the kilonova is bright enough. On the other hand, if the luminosity and timescale of energy injection from magnetar are lower and shorter, respectively, the light curve of kilonova emission only show one peak, and it can only be classified based on the color evolution in the later stage, but remain to need enough observational data to do that.

On the other hand, \cite{2022MNRAS.516.2614A} proposed to invoke the magnetar wind as a magnetized plasma blob injecting energy primarily through a forward shock, and ignore the contribution of termination shock to produce PWN. The efficiency of energy injection is dependent on optical depth, and it is very difficult to constrain \citep{2025ApJ...978...52A}. In this work, we consider the a termination shock which accelerates particles to produce non-thermal radiation, and the energy injection mechanism used in this paper is quite different from that of in \cite{2022MNRAS.516.2614A}. So that, we adopt a constant efficiency to do the calculations.

It is worth noting that the constant color evolution at the late state is not only caused by PWN, but is possible from the off-axis jet afterglow \citep{2018MNRAS.481.1597G,2018PTEP.2018d3E02I} and recombination \citep{2024A&A...688A..95S}. The light curve produced by an off-axis jet afterglow could potentially resemble to that of the light curve by considering three energy components with similar observational effects \citep{2005ApJ...631.1022G,2010ApJ...722..235V,2020MNRAS.498.5643T}, while the plotted color evolution exhibits a "rise-decline-stabilization" behavior . Those two effects cannot be ruled out only based on the constant color evolution of kilonova emission at the late state.

The space-based multi-band Astronomical Variable Objects Monitor (SVOM) mission, which is used to study the GRBs and those associated with SNe and kilonovae, was successfully launched on June 22, 2024 \citep{2016arXiv161006892W}. One of the main scientific goals is to follow-up observations of kilonova candidates via its visible-band telescope (VT) together with other optical survey projects after the short-duration GRBs trigger. Thus, it is expected that SVOM will be helpful in making deep observations of kilonova candidates in the future, and to study more details of the progenitors and central engines of short-duration GRBs.

\section*{Acknowledgements}

This work is supported by the Guangxi Science Foundation (grant Nos. 2023GXNSFDA026007 and 2025GXNSFDA02850010), the Natural Science Foundation of China (grant Nos. 12494574, 11922301 and 12133003), the Program of Bagui Scholars Program (LHJ), and the Guangxi Talent Program (“Highland of Innovation Talents”).

\section*{Data Availability}
This is a theoretical study that do not generate any new data associated with this article. If one needs to adopt the calculated data in this article, it should be to cite this reference paper.




\begin{thebibliography}{99}
\bibitem[\protect\citeauthoryear{Abbott et al.}{2017}]{2017PhRvL.119p1101A} Abbott B.~P., Abbott R., Abbott T.~D., Acernese F., Ackley K., Adams C., Adams T., et al., 2017, PhRvL, 119, 161101. doi:10.1103/PhysRevLett.119.161101
\bibitem[\protect\citeauthoryear{Abbott et al.}{2021}]{2021ApJ...915L...5A} Abbott R., Abbott T.~D., Abraham S., Acernese F., Ackley K., Adams A., Adams C., et al., 2021, ApJL, 915, L5. doi:10.3847/2041-8213/ac082e
\bibitem[\protect\citeauthoryear{Abbott et al.}{2023}]{2023PhRvX..13d1039A} Abbott R., Abbott T.~D., Acernese F., Ackley K., Adams C., Adhikari N., Adhikari R.~X., et al., 2023, PhRvX, 13, 041039. doi:10.1103/PhysRevX.13.041039
\bibitem[\protect\citeauthoryear{Ai, Zhang, \& Zhu}{2022}]{2022MNRAS.516.2614A} Ai S., Zhang B., Zhu Z., 2022, MNRAS, 516, 2614. doi:10.1093/mnras/stac2380
\bibitem[\protect\citeauthoryear{Ai, Gao, \& Zhang}{2025}]{2025ApJ...978...52A} Ai S., Gao H., Zhang B., 2025, ApJ, 978, 52. doi:10.3847/1538-4357/ad93b4
\bibitem[\protect\citeauthoryear{Barnes et al.}{2016}]{2016ApJ...829..110B} Barnes J., Kasen D., Wu M.-R., Mart{\'\i}nez-Pinedo G., 2016, ApJ, 829, 110. doi:10.3847/0004-637X/829/2/110
\bibitem[\protect\citeauthoryear{Berger}{2011}]{2011NewAR..55....1B} Berger E., 2011, NewAR, 55, 1. doi:10.1016/j.newar.2010.10.001
\bibitem[\protect\citeauthoryear{Berger}{2014}]{2014ARA&A..52...43B} Berger E., 2014, ARA\&A, 52, 43. doi:10.1146/annurev-astro-081913-035926
\bibitem[\protect\citeauthoryear{Chen et al.}{2017}]{2017ApJ...849..119C} Chen W., Xie W., Lei W.-H., Zou Y.-C., L{\"u} H.-J., Liang E.-W., Gao H., et al., 2017, ApJ, 849, 119. doi:10.3847/1538-4357/aa8f4a
\bibitem[\protect\citeauthoryear{Ciolfi et al.}{2017}]{2017PhRvD..95f3016C} Ciolfi R., Kastaun W., Giacomazzo B., Endrizzi A., Siegel D.~M., Perna R., 2017, PhRvD, 95, 063016. doi:10.1103/PhysRevD.95.063016
\bibitem[\protect\citeauthoryear{Collins et al.}{2023}]{2023MNRAS.521.1858C} Collins C.~E., Bauswein A., Sim S.~A., Vijayan V., Mart{\'\i}nez-Pinedo G., Just O., Shingles L.~J., et al., 2023, MNRAS, 521, 1858. doi:10.1093/mnras/stad606
\bibitem[\protect\citeauthoryear{Covino et al.}{2017}]{2017NatAs...1..791C} Covino S., Wiersema K., Fan Y.~Z., Toma K., Higgins A.~B., Melandri A., D'Avanzo P., et al., 2017, NatAs, 1, 791. doi:10.1038/s41550-017-0285-z
\bibitem[\protect\citeauthoryear{Dai \& Lu}{1998}]{1998PhRvL..81.4301D} Dai Z.~G., Lu T., 1998, PhRvL, 81, 4301. doi:10.1103/PhysRevLett.81.4301
\bibitem[\protect\citeauthoryear{Dai et al.}{2006}]{2006Sci...311.1127D} Dai Z.~G., Wang X.~Y., Wu X.~F., Zhang B., 2006, Sci, 311, 1127. doi:10.1126/science.1123606
\bibitem[\protect\citeauthoryear{Du et al.}{2025}]{2025ApJ...985...42D} Du Z.-W., L{\"u} H., Liu X., Fan X., Liang E., 2025, ApJ, 985, 42. doi:10.3847/1538-4357/adcb49
\bibitem[\protect\citeauthoryear{Eichler et al.}{1989}]{1989Natur.340..126E} Eichler D., Livio M., Piran T., Schramm D.~N., 1989, Natur, 340, 126. doi:10.1038/340126a0
\bibitem[\protect\citeauthoryear{Evans et al.}{2017}]{2017Sci...358.1565E} Evans P.~A., Cenko S.~B., Kennea J.~A., Emery S.~W.~K., Kuin N.~P.~M., Korobkin O., Wollaeger R.~T., et al., 2017, Sci, 358, 1565. doi:10.1126/science.aap9580
\bibitem[\protect\citeauthoryear{Fan \& Piran}{2006}]{2006MNRAS.369..197F} Fan Y., Piran T., 2006, MNRAS, 369, 197. doi:10.1111/j.1365-2966.2006.10280.x
\bibitem[\protect\citeauthoryear{Fan \& Xu}{2006}]{2006MNRAS.372L..19F} Fan Y.-Z., Xu D., 2006, MNRAS, 372, L19. doi:10.1111/j.1745-3933.2006.00217.x
\bibitem[\protect\citeauthoryear{Fan, Wu, \& Wei}{2013}]{2013PhRvD..88f7304F} Fan Y.-Z., Wu X.-F., Wei D.-M., 2013, PhRvD, 88, 067304. doi:10.1103/PhysRevD.88.067304
\bibitem[\protect\citeauthoryear{Gaensler \& Slane}{2006}]{2006ARA&A..44...17G} Gaensler B.~M., Slane P.~O., 2006, ARA\&A, 44, 17. doi:10.1146/annurev.astro.44.051905.092528
\bibitem[\protect\citeauthoryear{Gao, Zhang, \& L{\"u}}{2016}]{2016PhRvD..93d4065G} Gao H., Zhang B., L{\"u} H.-J., 2016, PhRvD, 93, 044065. doi:10.1103/PhysRevD.93.044065
\bibitem[\protect\citeauthoryear{Gao et al.}{2017}]{2017ApJ...837...50G} Gao H., Zhang B., L{\"u} H.-J., Li Y., 2017, ApJ, 837, 50. doi:10.3847/1538-4357/aa5be3
\bibitem[\protect\citeauthoryear{Giacomazzo \& Perna}{2013}]{2013ApJ...771L..26G} Giacomazzo B., Perna R., 2013, ApJL, 771, L26. doi:10.1088/2041-8205/771/2/L26
\bibitem[\protect\citeauthoryear{Goldstein et al.}{2017}]{2017ApJ...848L..14G} Goldstein A., Veres P., Burns E., Briggs M.~S., Hamburg R., Kocevski D., Wilson-Hodge C.~A., et al., 2017, ApJL, 848, L14. doi:10.3847/2041-8213/aa8f41
\bibitem[\protect\citeauthoryear{Gompertz et al.}{2018}]{2018ApJ...860...62G} Gompertz B.~P., Levan A.~J., Tanvir N.~R., Hjorth J., Covino S., Evans P.~A., Fruchter A.~S., et al., 2018, ApJ, 860, 62. doi:10.3847/1538-4357/aac206
\bibitem[\protect\citeauthoryear{Gottlieb et al.}{2025}]{2025ApJ...984...77G} Gottlieb O., Metzger B.~D., Foucart F., Ramirez-Ruiz E., 2025, ApJ, 984, 77. doi:10.3847/1538-4357/adc577
\bibitem[\protect\citeauthoryear{Granot}{2005}]{2005ApJ...631.1022G} Granot J., 2005, ApJ, 631, 1022. doi:10.1086/432676
\bibitem[\protect\citeauthoryear{Granot et al.}{2018}]{2018MNRAS.481.1597G} Granot J., Gill R., Guetta D., De Colle F., 2018, MNRAS, 481, 1597. doi:10.1093/mnras/sty2308
\bibitem[\protect\citeauthoryear{Hotokezaka et al.}{2013}]{2013PhRvD..87b4001H} Hotokezaka K., Kiuchi K., Kyutoku K., Okawa H., Sekiguchi Y.-. ichiro ., Shibata M., Taniguchi K., 2013, PhRvD, 87, 024001. doi:10.1103/PhysRevD.87.024001
\bibitem[\protect\citeauthoryear{Ioka \& Nakamura}{2018}]{2018PTEP.2018d3E02I} Ioka K., Nakamura T., 2018, PTEP, 2018, 043E02. doi:10.1093/ptep/pty036
\bibitem[\protect\citeauthoryear{Jin et al.}{2015}]{2015ApJ...811L..22J} Jin Z.-P., Li X., Cano Z., Covino S., Fan Y.-Z., Wei D.-M., 2015, ApJL, 811, L22. doi:10.1088/2041-8205/811/2/L22
\bibitem[\protect\citeauthoryear{Kasen \& Bildsten}{2010}]{2010ApJ...717..245K} Kasen D., Bildsten L., 2010, ApJ, 717, 245. doi:10.1088/0004-637X/717/1/245
\bibitem[\protect\citeauthoryear{Kasen et al.}{2017}]{2017Natur.551...80K} Kasen D., Metzger B., Barnes J., Quataert E., Ramirez-Ruiz E., 2017, Natur, 551, 80. doi:10.1038/nature24453
\bibitem[\protect\citeauthoryear{Kawaguchi, Shibata, \& Tanaka}{2020}]{2020ApJ...889..171K} Kawaguchi K., Shibata M., Tanaka M., 2020, ApJ, 889, 171. doi:10.3847/1538-4357/ab61f6
\bibitem[\protect\citeauthoryear{Korobkin et al.}{2012}]{2012MNRAS.426.1940K} Korobkin O., Rosswog S., Arcones A., Winteler C., 2012, MNRAS, 426, 1940. doi:10.1111/j.1365-2966.2012.21859.x
\bibitem[\protect\citeauthoryear{Liu et al.}{2025}]{2025ApJ...988L..46L} Liu X.-X., L{\"u} H.-J., Chen Q.-H., Du Z.-W., Liang E.-W., 2025, ApJL, 988, L46. doi:10.3847/2041-8213/adec83
\bibitem[\protect\citeauthoryear{L{\"u} et al.}{2015}]{2015ApJ...805...89L} L{\"u} H.-J., Zhang B., Lei W.-H., Li Y., Lasky P.~D., 2015, ApJ, 805, 89. doi:10.1088/0004-637X/805/2/89
\bibitem[\protect\citeauthoryear{L{\"u} et al.}{2017}]{2017ApJ...835..181L} L{\"u} H.-J., Zhang H.-M., Zhong S.-Q., Hou S.-J., Sun H., Rice J., Liang E.-W., 2017, ApJ, 835, 181. doi:10.3847/1538-4357/835/2/181
\bibitem[\protect\citeauthoryear{L{\"u} et al.}{2018}]{2018MNRAS.480.4402L} L{\"u} H.-J., Zou L., Lan L., Liang E.-W., 2018, MNRAS, 480, 4402. doi:10.1093/mnras/sty2176
\bibitem[\protect\citeauthoryear{L{\"u} et al.}{2019}]{2019MNRAS.486.4479L} L{\"u} H.-J., Shen J., Lan L., Rice J., Lei W.-H., Liang E.-W., 2019, MNRAS, 486, 4479. doi:10.1093/mnras/stz1155
\bibitem[\protect\citeauthoryear{L{\"u} et al.}{2020}]{2020ApJ...898L...6L} L{\"u} H.-J., Yuan Y., Lan L., Zhang B.-B., Zou J.-H., Peng Z.-K., Shen J., et al., 2020, ApJL, 898, L6. doi:10.3847/2041-8213/aba1ed
\bibitem[\protect\citeauthoryear{Lasky et al.}{2014}]{2014PhRvD..89d7302L} Lasky P.~D., Haskell B., Ravi V., Howell E.~J., Coward D.~M., 2014, PhRvD, 89, 047302. doi:10.1103/PhysRevD.89.047302
\bibitem[\protect\citeauthoryear{Li \& Paczy{\'n}ski}{1998}]{1998ApJ...507L..59L} Li L.-X., Paczy{\'n}ski B., 1998, ApJL, 507, L59. doi:10.1086/311680
\bibitem[\protect\citeauthoryear{Ma et al.}{2018}]{2018ApJ...852L...5M} Ma S.-B., Lei W.-H., Gao H., Xie W., Chen W., Zhang B., Wang D.-X., 2018, ApJL, 852, L5. doi:10.3847/2041-8213/aaa0cd
\bibitem[\protect\citeauthoryear{Metzger, Quataert, \& Thompson}{2008}]{2008MNRAS.385.1455M} Metzger B.~D., Quataert E., Thompson T.~A., 2008, MNRAS, 385, 1455. doi:10.1111/j.1365-2966.2008.12923.x
\bibitem[\protect\citeauthoryear{Metzger et al.}{2010}]{2010MNRAS.406.2650M} Metzger B.~D., Mart{\'\i}nez-Pinedo G., Darbha S., Quataert E., Arcones A., Kasen D., Thomas R., et al., 2010, MNRAS, 406, 2650. doi:10.1111/j.1365-2966.2010.16864.x
\bibitem[\protect\citeauthoryear{Metzger \& Piro}{2014}]{2014MNRAS.439.3916M} Metzger B.~D., Piro A.~L., 2014, MNRAS, 439, 3916. doi:10.1093/mnras/stu247
\bibitem[\protect\citeauthoryear{Metzger}{2017}]{2017LRR....20....3M} Metzger B.~D., 2017, LRR, 20, 3. doi:10.1007/s41114-017-0006-z
\bibitem[\protect\citeauthoryear{Metzger}{2019}]{2019LRR....23....1M} Metzger B.~D., 2019, LRR, 23, 1. doi:10.1007/s41114-019-0024-0
\bibitem[\protect\citeauthoryear{Murase et al.}{2015}]{2015ApJ...805...82M} Murase K., Kashiyama K., Kiuchi K., Bartos I., 2015, ApJ, 805, 82. doi:10.1088/0004-637X/805/1/82
\bibitem[\protect\citeauthoryear{Murase, Kashiyama, \& M{\'e}sz{\'a}ros}{2016}]{2016MNRAS.461.1498M} Murase K., Kashiyama K., M{\'e}sz{\'a}ros P., 2016, MNRAS, 461, 1498. doi:10.1093/mnras/stw1328
\bibitem[\protect\citeauthoryear{Nagakura et al.}{2014}]{2014ApJ...784L..28N} Nagakura H., Hotokezaka K., Sekiguchi Y., Shibata M., Ioka K., 2014, ApJL, 784, L28. doi:10.1088/2041-8205/784/2/L28
\bibitem[\protect\citeauthoryear{Nakar}{2020}]{2020PhR...886....1N} Nakar E., 2020, PhR, 886, 1. doi:10.1016/j.physrep.2020.08.008
\bibitem[\protect\citeauthoryear{Nollert}{1999}]{1999CQGra..16R.159N} Nollert H.-P., 1999, CQGra, 16, R159. doi:10.1088/0264-9381/16/12/201
\bibitem[\protect\citeauthoryear{Paczynski}{1986}]{1986ApJ...308L..43P} Paczynski B., 1986, ApJL, 308, L43. doi:10.1086/184740
\bibitem[\protect\citeauthoryear{Paczynski}{1991}]{1991AcA....41..257P} Paczynski B., 1991, AcA, 41, 257
\bibitem[\protect\citeauthoryear{Piro et al.}{2019}]{2019MNRAS.483.1912P} Piro L., Troja E., Zhang B., Ryan G., van Eerten H., Ricci R., Wieringa M.~H., et al., 2019, MNRAS, 483, 1912. doi:10.1093/mnras/sty3047
\bibitem[\protect\citeauthoryear{Ravi \& Lasky}{2014}]{2014MNRAS.441.2433R} Ravi V., Lasky P.~D., 2014, MNRAS, 441, 2433. doi:10.1093/mnras/stu720
\bibitem[\protect\citeauthoryear{Ren et al.}{2019}]{2019ApJ...885...60R} Ren J., Lin D.-B., Zhang L.-L., Li X.-Y., Liu T., Lu R.-J., Wang X.-G., et al., 2019, ApJ, 885, 60. doi:10.3847/1538-4357/ab4188
\bibitem[\protect\citeauthoryear{Rezzolla et al.}{2010}]{2010CQGra..27k4105R} Rezzolla L., Baiotti L., Giacomazzo B., Link D., Font J.~A., 2010, CQGra, 27, 114105. doi:10.1088/0264-9381/27/11/114105
\bibitem[\protect\citeauthoryear{Rezzolla et al.}{2011}]{2011ApJ...732L...6R} Rezzolla L., Giacomazzo B., Baiotti L., Granot J., Kouveliotou C., Aloy M.~A., 2011, ApJL, 732, L6. doi:10.1088/2041-8205/732/1/L6
\bibitem[\protect\citeauthoryear{Rosswog et al.}{2000}]{2000A&A...360..171R} Rosswog S., Davies M.~B., Thielemann F.-K., Piran T., 2000, A\&A, 360, 171. doi:10.48550/arXiv.astro-ph/0005550
\bibitem[\protect\citeauthoryear{Rosswog, Piran, \& Nakar}{2013}]{2013MNRAS.430.2585R} Rosswog S., Piran T., Nakar E., 2013, MNRAS, 430, 2585. doi:10.1093/mnras/sts708
\bibitem[\protect\citeauthoryear{Rosswog et al.}{2014}]{2014MNRAS.439..744R} Rosswog S., Korobkin O., Arcones A., Thielemann F.-K., Piran T., 2014, MNRAS, 439, 744. doi:10.1093/mnras/stt2502
\bibitem[\protect\citeauthoryear{Rowlinson et al.}{2013}]{2013MNRAS.430.1061R} Rowlinson A., O'Brien P.~T., Metzger B.~D., Tanvir N.~R., Levan A.~J., 2013, MNRAS, 430, 1061. doi:10.1093/mnras/sts683
\bibitem[\protect\citeauthoryear{Sari, Piran, \& Narayan}{1998}]{1998ApJ...497L..17S} Sari R., Piran T., Narayan R., 1998, ApJL, 497, L17. doi:10.1086/311269
\bibitem[\protect\citeauthoryear{Savchenko et al.}{2017}]{2017ApJ...848L..15S} Savchenko V., Ferrigno C., Kuulkers E., Bazzano A., Bozzo E., Brandt S., Chenevez J., et al., 2017, ApJL, 848, L15. doi:10.3847/2041-8213/aa8f94
\bibitem[\protect\citeauthoryear{Siegel \& Metzger}{2018}]{2018ApJ...858...52S} Siegel D.~M., Metzger B.~D., 2018, ApJ, 858, 52. doi:10.3847/1538-4357/aabaec
\bibitem[\protect\citeauthoryear{{\v{S}}imon et al.}{2001}]{2001A&A...377..450S} {\v{S}}imon V., Hudec R., Pizzichini G., Masetti N., 2001, A\&A, 377, 450. doi:10.1051/0004-6361:20011158
\bibitem[\protect\citeauthoryear{{\v{S}}imon, Hudec, \& Pizzichini}{2004}]{2004A&A...427..901S} {\v{S}}imon V., Hudec R., Pizzichini G., 2004, A\&A, 427, 901. doi:10.1051/0004-6361:20041548
\bibitem[\protect\citeauthoryear{Sneppen et al.}{2024}]{2024A&A...688A..95S} Sneppen A., Watson D., Gillanders J.~H., Heintz K.~E., 2024, A\&A, 688, A95. doi:10.1051/0004-6361/202348758
\bibitem[\protect\citeauthoryear{Tanaka \& Takahara}{2010}]{2010ApJ...715.1248T} Tanaka S.~J., Takahara F., 2010, ApJ, 715, 1248. doi:10.1088/0004-637X/715/2/1248
\bibitem[\protect\citeauthoryear{Tanaka \& Takahara}{2013}]{2013MNRAS.429.2945T} Tanaka S.~J., Takahara F., 2013, MNRAS, 429, 2945. doi:10.1093/mnras/sts528
\bibitem[\protect\citeauthoryear{Troja et al.}{2020}]{2020MNRAS.498.5643T} Troja E., van Eerten H., Zhang B., Ryan G., Piro L., Ricci R., O'Connor B., et al., 2020, MNRAS, 498, 5643. doi:10.1093/mnras/staa2626
\bibitem[\protect\citeauthoryear{Troja}{2023}]{2023Univ....9..245T} Troja E., 2023, Univ, 9, 245. doi:10.3390/universe9060245
\bibitem[\protect\citeauthoryear{van Eerten, Zhang, \& MacFadyen}{2010}]{2010ApJ...722..235V} van Eerten H., Zhang W., MacFadyen A., 2010, ApJ, 722, 235. doi:10.1088/0004-637X/722/1/235
\bibitem[\protect\citeauthoryear{Wang et al.}{2024}]{2024ApJ...963..156W} Wang S.-N., L{\"u} H.-J., Yuan Y., Yuan H.-Y., Rice J., Chen M.-H., Liang E.-W., 2024, ApJ, 963, 156. doi:10.3847/1538-4357/ad2205
\bibitem[\protect\citeauthoryear{Wei et al.}{2016}]{2016arXiv161006892W} Wei J., Cordier B., Antier S., Antilogus P., Atteia J.-L., Bajat A., Basa S., et al., 2016, arXiv, arXiv:1610.06892. doi:10.48550/arXiv.1610.06892
\bibitem[\protect\citeauthoryear{Yang et al.}{2015}]{2015NatCo...6.7323Y} Yang B., Jin Z.-P., Li X., Covino S., Zheng X.-Z., Hotokezaka K., Fan Y.-Z., et al., 2015, NatCo, 6, 7323. doi:10.1038/ncomms8323
\bibitem[\protect\citeauthoryear{Yu, Zhang, \& Gao}{2013}]{2013ApJ...776L..40Y} Yu Y.-W., Zhang B., Gao H., 2013, ApJL, 776, L40. doi:10.1088/2041-8205/776/2/L40
\bibitem[\protect\citeauthoryear{Yu, Liu, \& Dai}{2018}]{2018ApJ...861..114Y} Yu Y.-W., Liu L.-D., Dai Z.-G., 2018, ApJ, 861, 114. doi:10.3847/1538-4357/aac6e5
\bibitem[\protect\citeauthoryear{Yuan et al.}{2021}]{2021ApJ...912...14Y} Yuan Y., L{\"u} H.-J., Yuan H.-Y., Ma S.-B., Lei W.-H., Liang E.-W., 2021, ApJ, 912, 14. doi:10.3847/1538-4357/abedb1
\bibitem[\protect\citeauthoryear{Zhang \& M{\'e}sz{\'a}ros}{2001}]{2001ApJ...552L..35Z} Zhang B., M{\'e}sz{\'a}ros P., 2001, ApJL, 552, L35. doi:10.1086/320255
\bibitem[\protect\citeauthoryear{Zhang}{2011}]{2011CRPhy..12..206Z} Zhang B., 2011, CRPhy, 12, 206. doi:10.1016/j.crhy.2011.03.004
\bibitem[\protect\citeauthoryear{Zhang}{2013}]{2013ApJ...763L..22Z} Zhang B., 2013, ApJL, 763, L22. doi:10.1088/2041-8205/763/1/L22
\bibitem[\protect\citeauthoryear{Zhang et al.}{2018}]{2018NatCo...9..447Z} Zhang B.-B., Zhang B., Sun H., Lei W.-H., Gao H., Li Y., Shao L., et al., 2018, NatCo, 9, 447. doi:10.1038/s41467-018-02847-3


\end{thebibliography}



\bsp	
\label{lastpage}
\end{document}